\documentclass[%
twocolumn,
preprintnumbers,
 amsmath,amssymb,
prc,
]{revtex4-2}

\usepackage{graphicx}
\usepackage{dcolumn}
\usepackage{bm}

\usepackage{aas_macros}
\usepackage{color}
\usepackage{amsmath}

\usepackage{setspace}
\usepackage{subfigure}

\newcommand{\average}[1]{\ensuremath{\langle#1\rangle}}

\usepackage{hyperref}
\hypersetup{
    colorlinks=true,
    linkcolor=blue,
    citecolor=blue,        
    filecolor=magenta,      
    urlcolor=blue,
}

\usepackage{braket}

\begin{document}

\preprint{LA-UR-22-32329}
\title{QRPA calculations for M1 transitions with the noniterative finite amplitude method and the application to neutron radiative capture cross sections
}



\author{Hirokazu Sasaki}
\email{hsasaki@lanl.gov}

\affiliation{Theoretical Division, Los Alamos National Laboratory, Los Alamos, New Mexico 87545, USA}
\author{Toshihiko Kawano}

\affiliation{Theoretical Division, Los Alamos National Laboratory, Los Alamos, New Mexico 87545, USA}
\author{Ionel Stetcu}

\affiliation{Theoretical Division, Los Alamos National Laboratory, Los Alamos, New Mexico 87545, USA}




\date{\today}

\begin{abstract}

We derive the equations of quasiparticle random-phase approximation (QRPA) based on the finite amplitude method (FAM) with the Hartree-Fock+BCS (HF+BCS) single-particle states, and calculate the magnetic dipole (M1) transition for deformed gadolinium isotopes. Our QRPA calculation shows both large spin-flip transitions in the 5 to 10 MeV excitation energy and the low energy orbital transition that would correspond to the M1 scissors mode observed in nuclear experiments. Then, we calculate neutron capture reactions based on the statistical Hauser-Feshbach theory with the photoabsorption cross sections given by QRPA. We find that the capture cross section is enhanced due to the contribution from the low energy M1 transition although the calculated capture cross section still underestimates the experimental data. This issue in the calculated capture cross section could be improved by uncertainties of low energy E1 transition neglected in our QRPA calculation.
\end{abstract}

\maketitle


\section{Introduction}
Neutron-capture reactions are essential for the nucleosynthesis of heavy elements, and their cross section data are needed to calculate nuclear abundances in astrophysical sites \cite{Burbidge:1957vc}. It is theoretically predicted that many unstable nuclei are produced through the rapid neutron capture process ($r$ process) \cite{Mumpower:2015ova} before $\beta$-decays in core-collapse supernovae and neutron star mergers, which contributes to the galactic chemical evolution of about half of heavy elements \cite{Kobayashi:2020jes}. The $(n,\gamma)$ reaction is also important for neutrino-induced nucleosynthesis inside core-collapse supernovae such as the $\nu p$ process \cite{Frohlich:2005ys,Sasaki:2017jry,Sasaki:2021ffa} and the $\nu$ process \cite{Woosley:1989bd,Hayakawa:2018ekx,Ko:2022uqv}. The experimental data for the capture reactions are mainly limited to stable nuclei, so a reliable theoretical calculation is required to study the origin of elements. 

In the neutron capture reaction, a compound nucleus is formed after the interaction of an incoming neutron with the target nucleus and finally decays by emitting several $\gamma$-rays. The statistical Hauser-Feshbach theory \cite{Hauser:1952zz} can estimate the decay rates of such a compound nucleus, and calculate neutron capture cross sections with transmission coefficients for all possible competing channels. The transmission coefficient of the outgoing $\gamma$-ray is calculated with $\gamma$-ray strength functions \cite{Kopecky:1990PRC} of electric and magnetic giant resonances under the Brink-Axel hypothesis \cite{Brink:1955phd}. Instead of a standard Lorentzian using the giant dipole resonance (GDR) parameters, a generalized Lorentzian \cite{Kopecky:1990PRC,Kopecky:1993zz} is proposed empirically and widely used to calculate the $\gamma$-ray strength function of the electric dipole (E1) transition. The magnetic dipole (M1) transition also has non-negligible impact on the capture cross section. In particular, the M1 scissors mode, which is often observed in a few MeV energy region for strongly deformed nuclei, can enhance the capture cross section \cite{Ullmann:2014roa,Mumpower:2017gqj,Goriely:2018gig}. 

The M1 scissors mode is induced by the collective motion of protons and neutrons inside a deformed nucleus associated with the orbital angular momentum operator, and it was first observed in an electron scattering experiment \cite{Bohle:1983sx}. Since then, the M1 scissors mode has been found in various rare-earth nuclei and actinides \cite{Heyde:2010ng}. 

Calculating capture reactions without any phenomenological parameters as used in the conventional $\gamma$-ray strength function is possible when we employ the strength of the M1 transition obtained in the density functional theory (DFT) \cite{ring2004nuclear}. DFT interprets the giant resonance as coherent superpositions of 1$p$-1$h$ excitations in a nuclear many-body system, and the transition strength of such a collective excitation is microscopically calculated with random-phase approximation (RPA) \cite{ring2004nuclear}. In order to introduce pairing correlations, essential for description of open-shell nuclei, RPA must be extended to quasiparticle RPA (QRPA) \cite{rowe2010nuclear}. Various collective excitations have been calculated in (Q)RPA \cite{harakeh2001giant}. Fully self-consistent (Q)RPA approaches with the Skyrme forces have been applied to the E1 transition \cite{Yoshida:2010zu,Inakura:2009vs,Oishi:2015lph,Sasaki:2022ipn} as well as the M1 transition \cite{Vesely:2009eb,Nesterenko:2010ra,Repko:2018gcn,Tselyaev:2019ovd,Nesterenko:2021wkm,Nesterenko:2022a}. Although the (Q)RPA for the E1 transition well reproduces the resonance energy of GDR in heavy nuclei, it is often reported that the calculated spin-flip M1 giant resonance unsatisfactory agrees with experimental data and depends on parameterizations of Skyrme force \cite{Nesterenko:2010ra}. The spin densities in the Skyrme force increase the resonance peak of the spin-flip transition \cite{Vesely:2009eb}.

The finite amplitude method (FAM) \cite{Nakatsukasa:2007qj,Avogadro:2011gd} is an efficient way to solve the fully self-consistent (Q)RPA equation and applied to various multipole collective excitations \cite{Inakura:2009vs,Stoitsov:2011zz,Hinohara:2013qda,Kortelainen:2015gxa,Oishi:2015lph,Sasaki:2022ipn}. The FAM approach was extended to the relativistic framework \cite{Liang:2013pda,Niksic:2013ega}, the proton-neutron FAM (pnFAM) for weak interactions \cite{Mustonen:2014bya,Shafer:2016etk,Ney:2020mnx,Hinohara:2022uip}, and the fission calculation \cite{Washiyama:2020qfr}. The framework of FAM-RPA is used to derive the RPA matrices for Skyrme functionals by the explicit linearization of residual interactions \cite{Sasaki:2022ipn}. The technique of such noniterative FAM-RPA can be used to derive QRPA matrices based on the framework of FAM-QRPA that enables the calculation of the M1 transition for open-shell and deformed nuclei without any iterative procedure to obtain forward and backward amplitudes in other conventional FAM-QRPA.

In this paper, we derive the QRPA matrices based on noniterative FAM-QRPA and calculate the M1 transition for deformed gadolinium isotopes. The microscopically calulated photoabsorption cross sections for the E1 and M1 transitions with our QRPA approach are then fed into the Hauser-Feshbach calculations as the $\gamma$-ray transmission coefficients to obtain the neutron capture cross sections. Then, we compare the calculated result with available experimental data.

\section{Theory}

\label{sec:theory}

\subsection{Finite amplitude method (FAM)}
We briefly review the general framework of the finite amplitude method in QRPA. In FAM-QRPA, the forward and backward amplitudes of a frequency $\omega$ are calculated through \cite{Avogadro:2011gd,Oishi:2015lph},
\begin{eqnarray}
    (E_{\mu}+E_{\nu}-\omega)X_{\mu\nu}(\omega)+\delta H^{20}_{\mu\nu}(\omega)&=&-F^{20}_{\mu\nu}(\omega), \label{eq:FAM-QRPA eq x}\\
    (E_{\mu}+E_{\nu}+\omega)Y_{\mu\nu}(\omega)+\delta H^{02}_{\mu\nu}(\omega)&=&-F^{02}_{\mu\nu}(\omega), \label{eq:FAM-QRPA eq y}
\end{eqnarray}
where $X(Y)$ is the forward (backward) amplitude, $E_{\mu}$ is an energy eigenvalue of a Bogoliubov quasiparticle state $\mu$, $\delta H^{20(02)}_{\mu\nu}(\omega)$ is the two-quasiparticle component of residual interactions, and $F^{20(02)}_{\mu\nu}(\omega)$ is the two-quasiparticle component of the external field. For a simple notation, hereafter, we drop the index $\omega$ in matrices except when necessary. The residual interaction is calculated with the forward and backward amplitudes, and these amplitudes are determined by solving Eqs.~(\ref{eq:FAM-QRPA eq x}) and (\ref{eq:FAM-QRPA eq y}) iteratively at each $\omega$. The obtained amplitudes are used to calculate the transition strength \cite{Avogadro:2011gd},
\begin{equation}
    \frac{\mathrm{d}B(\omega;F)}{\mathrm{d}\omega}=-\frac{1}{2\pi}\mathrm{Im}\sum_{\mu\nu}(F_{\mu\nu}^{20*}X_{\mu\nu}+F_{\mu\nu}^{02*}Y_{\mu\nu}
    ),\label{eq:transition strength QRPA}
\end{equation}
where the imaginary part of $\omega$ corresponds to the Lorentzian width characterizing the width of the transition strength. Hereafter, we assume that the external field $F$ does not change the isospin of nucleons.

\subsection{QRPA equations}
\label{sec:QRPA equation}
Instead of solving Eqs.~(\ref{eq:FAM-QRPA eq x}) and (\ref{eq:FAM-QRPA eq y}) by an iterative procedure, we derive the QRPA equation with the well-known QRPA matrices $A$ and $B$ from the explicit linearization of the residual interaction in Eqs.(\ref{eq:FAM-QRPA eq x}) and (\ref{eq:FAM-QRPA eq y}) as in FAM-RPA \cite{Sasaki:2022ipn}. We use the quasiparticle states of Hamiltonian in the Hartree-Fock+BCS (HF+BCS) calculation instead of the Hartree-Fock-Bogoliubov (HFB) calculation. In HF+BCS, the HFB matrices such as $U$ and $V$ only allow the mixing between a single-particle state $k$, and the corresponding time-reversed state $\bar{k}$. These matrices can be described in the coordinate space $\vec{r}$, the spin space $\sigma$, and the isospin $q (=n,p)$ \cite{ring2004nuclear,rowe2010nuclear},
\begin{eqnarray}
    U_{\mu}(\vec{r},\sigma,q)&=&u_{\mu}\phi_{\mu}(\vec{r},\sigma,q), \label{eq:u mu}\\
    V_{\mu}(\vec{r},\sigma,q)&=&-v_{\mu}\phi^{*}_{\bar{\mu}}(\vec{r},\sigma,q) \label{eq:v mu},
\end{eqnarray}
where $u_{\mu}\geq0$ and $v_{\mu}\geq0$ are the BCS parameters, and $\phi_{\mu}(\vec{r},\sigma,q)\equiv\phi^{q}_{\mu}$ is the single-particle state of the HF Hamiltonian $h_{0}$ satisfying $h_{0}\phi^{q}_{\mu}=\epsilon_{\mu}\phi^{q}_{\mu}$. To derive Eq.~(\ref{eq:v mu}), we use properties of the time-reversal symmetry of the HF single-particle states, $\phi^{q}_{\bar{\mu}}=T\phi^{q}_{\mu}$, and $\phi^{q}_{\bar{\bar{\mu}}}=T^{2}\phi^{q}_{\mu}=-\phi^{q}_{\mu}$ where $T$ is the time-reversal operator \cite{Vautherin:1973zz}. In the case of a Hermitian one-body external field $F$ \cite{Avogadro:2011gd,ring2004nuclear}, the quasiparticle components in Eqs.~(\ref{eq:FAM-QRPA eq x}) and (\ref{eq:FAM-QRPA eq y}) are given by
\begin{eqnarray}
    F^{20}_{\mu\nu}&=&\left(
    U^{\dagger}f V^{*}-V^{\dagger}f^{T}U^{*}
    \right)_{\mu\nu} \nonumber \\
    &=&-u_{\mu}v_{\nu}f_{\mu\bar{\nu}}^{q}+u_{\nu}v_{\mu}f_{\nu\bar{\mu}}^{q} \nonumber \\
    &=&-\zeta^{\tau}_{\mu\nu}f_{\mu\bar{\nu}}^{q}, \label{eq:F20}
\end{eqnarray}
\begin{eqnarray}
    F^{02}_{\mu\nu}&=&\left(
    U^{T}f^{T} V-V^{T}fU
    \right)_{\mu\nu} \nonumber\\
    &=&-u_{\mu}v_{\nu}f_{\bar{\nu}\mu}^{q}+u_{\nu}v_{\mu}f_{\bar{\mu}\nu}^{q} \nonumber\\
    &=&-\zeta^{\tau}_{\mu\nu}f_{\bar{\nu}\mu}^{q},\label{eq:F02}
\end{eqnarray}
where $\zeta^{\tau}_{\mu\nu}=u_{\mu}v_{\nu}+\tau u_{\nu}v_{\mu}$ with $\tau=\pm1$ and $f_{\mu\nu}^{q}=\int\mathrm{d}^{3}r\phi^{q*}_{\mu}F\phi^{q}_{\nu}$. From the second lines to the third lines in the above equations, we assume the time-reversal invariance for the external field, $TFT^{-1}=\tau F$, and use a relation, $f_{\nu\bar{\mu}}^{q}=-\tau f_{\mu\bar{\nu}}^{q}$ \cite{rowe2010nuclear,greiner1996nuclear}. 

We ignore the contributions from the residual interactions of the BCS pairing gap $\delta\Delta^{\pm}$ and the abnormal density $\delta \kappa^{\pm}$  \cite{Avogadro:2011gd}. Then, as in Eqs.~(\ref{eq:F20}) and (\ref{eq:F02}), quasiparticle components of the residual interaction in Eqs.~(\ref{eq:FAM-QRPA eq x}) and (\ref{eq:FAM-QRPA eq y}) are given by
\begin{eqnarray}
    \delta H^{20}_{\mu\nu}=-\zeta^{+}_{\mu\nu}\delta h_{\mu\bar{\nu}}^{\rm{even}}-\zeta^{-}_{\mu\nu}\delta h_{\mu\bar{\nu}}^{\rm{odd}}, \label{eq:dH20}\\
    \delta H^{02}_{\mu\nu}=-\zeta^{+}_{\mu\nu}\delta h_{\bar{\nu}\mu}^{\rm{even}}-\zeta^{-}_{\mu\nu}\delta h_{\bar{\nu}\mu}^{\rm{odd}}, \label{eq:dH02}
\end{eqnarray}
where $\delta h^{\rm{even(odd)}}$ is the residual interaction composed of time-even (odd) fields composed of the HF+BCS single-particle states. In FAM-QRPA, these residual interactions are calculated with a small parameter $\eta$ following the same procedure to obtain $\delta h(\omega)$ and $\delta\rho(\omega)$ in Ref.~\cite{Avogadro:2011gd}. In the limit of $\eta\to0$, such calculated residual interactions can be expressed as the linear combination of $X$ and $Y$ as done in FAM-RPA \cite{Sasaki:2022ipn}. Then, the QRPA equation is derived from Eqs.~(\ref{eq:FAM-QRPA eq x}) and (\ref{eq:FAM-QRPA eq y}), 
\begin{equation}
\left(
\begin{array}{cc}
     A -\omega& B \\
     B^{*} & A^{*}+\omega 
\end{array}
\right)
\left(
\begin{array}{c}
     X_{\alpha\beta}^{q^{\prime}}\\
     Y_{\alpha\beta}^{q^{\prime}}
\end{array}
\right)
=-\zeta^{\tau}_{\mu\nu}
\left(
\begin{array}{c}
     f_{\mu\nu}^{q}\\
     f_{\nu\mu}^{q}
\end{array}
\right),
\label{eq:QRPAeq}
\end{equation}

\begin{eqnarray}
A^{q,q^{\prime}}_{\mu\nu,\alpha\beta}&=&\left(E_{\mu}+E_{\nu}
    \right)\delta_{\mu\alpha}\delta_{\nu\beta} \nonumber\\
    &+&\zeta^{+}_{\mu\nu}\zeta^{+}_{\alpha\beta}\int\mathrm{d}^{3}r\ \phi_{\mu}^{q*}
      \left.
        \frac{\partial h_{q}^{\rm{even}}}{\partial (\eta\zeta^{+}_{\alpha\beta} X_{\alpha\beta}^{q^{\prime}})} 
      \right|_{\eta=0}\phi_{\nu}^{q} \nonumber\\
    &+&\zeta^{-}_{\mu\nu}\zeta^{-}_{\alpha\beta}\int\mathrm{d}^{3}r\ \phi_{\mu}^{q*}
      \left.
        \frac{\partial h_{q}^{\rm{odd}}}{\partial (\eta \zeta^{-}_{\alpha\beta}X_{\alpha\beta}^{q^{\prime}})} 
      \right|_{\eta=0}\phi_{\nu}^{q}, 
      \label{eq:QRPA matrix A}
\end{eqnarray}

\begin{eqnarray}
B^{q,q^{\prime}}_{\mu\nu,\alpha\beta}&=& \zeta^{+}_{\mu\nu}\zeta^{+}_{\alpha\beta}\int\mathrm{d}^{3}r\ \phi_{\mu}^{q*}
      \left.
        \frac{\partial h_{q}^{\rm{even}}}{\partial (\eta\zeta^{+}_{\alpha\beta} Y_{\alpha\beta}^{q^{\prime}})} 
      \right|_{\eta=0}\phi_{\nu}^{q}  \nonumber\\
    &+& \zeta^{-}_{\mu\nu}\zeta^{-}_{\alpha\beta}\int\mathrm{d}^{3}r\ \phi_{\mu}^{q*}
      \left.
        \frac{\partial h_{q}^{\rm{odd}}}{\partial (\eta\zeta^{-}_{\alpha\beta} Y_{\alpha\beta}^{q^{\prime}})}
      \right|_{\eta=0}\phi_{\nu}^{q} , \label{eq:QRPA matrix B}
\end{eqnarray}

\begin{eqnarray}
E_{\mu}=\sqrt{(\epsilon_{\mu}-\lambda)^{2}+\Delta^{2}_{\mu}},    \label{eq:BCS energy}
\end{eqnarray}
where $h_{q}$ is the time-dependent Hartree-Fock (TDHF) Hamiltonian of the nucleon $q$, and $h_{q}^{\mathrm{even(odd)}}$ is the time-even(odd) part of $h_{q}$, $\lambda$ is the Fermi energy, and $\Delta_{\mu}$ is the pairing gap of single-particle state $\mu$, and the size of the configuration space is restricted to $\mu\geq\nu,$ $\alpha\geq\beta$. When we derive Eqs.~(\ref{eq:QRPAeq})--(\ref{eq:BCS energy}), we assume the time-reversal symmetry of the HF Hamiltonian, $Th_{0}T^{-1}=h_{0}$, and change the definition of the amplitudes in Eqs.~(\ref{eq:FAM-QRPA eq x}) and (\ref{eq:FAM-QRPA eq y}) as 
\begin{eqnarray}
X_{\mu\nu}\to-X_{\mu\bar{\nu}},\ Y_{\mu\nu}\to-Y_{\mu\bar{\nu}}, \label{eq:xytrans}
\end{eqnarray}
which facilitates to extend an existing RPA code to the full QRPA calculation. The QRPA equation of Eq.~(\ref{eq:QRPAeq}) reproduces the RPA equation in Ref.~\cite{Sasaki:2022ipn} when we impose $\epsilon_{\mu(\alpha)}>\lambda>\epsilon_{\nu(\beta)}$ and $u_{\mu(\alpha)}=v_{\nu(\beta)}=1$, which results in $\zeta^{\pm}_{\mu\nu}=\zeta^{\pm}_{\alpha\beta}=1$ in Eqs.~(\ref{eq:QRPAeq})--(\ref{eq:QRPA matrix B}) and $\Delta_{\mu}=0$ in Eq.~(\ref{eq:BCS energy}). 

The integrands in Eqs.~(\ref{eq:QRPA matrix A}) and (\ref{eq:QRPA matrix B}) are calculated with the Skyrme forces composed of the HF+BCS single-particle states. For example, the contribution from the effective mass $m_{q}^{*}$ \cite{Maruhn:2013mpa} to the integral term proportional to $\zeta^{+}_{\mu\nu}\zeta^{+}_{\alpha\beta}$ in Eq.~(\ref{eq:QRPA matrix A}) is given by
\begin{align}
\label{eq:derivative of effective mass}
&\int\mathrm{d}^{3}r\ \phi_{\mu}^{q*}\nabla
\cdot\left(\frac{\partial}{\partial (\eta \zeta^{+}_{\alpha\beta}X_{\alpha\beta}^{q^{\prime}})}
\frac{-\hbar^{2}}{2m^{*}_{q}}\right)_{\eta=0}\nabla\phi_{\nu}^{q} \notag \\
&=\int\mathrm{d}^{3}r\
\left\{\frac{\partial(b_{1}\rho_{n}+b_{1}\rho_{p}-b_{1}^{\prime}\rho_{q})}{\partial (\eta \zeta^{+}_{\alpha\beta}X_{\alpha\beta}^{q^{\prime}})}
\right\}_{\eta=0}\nabla\phi_{\mu}^{q*}\cdot\nabla\phi_{\nu}^{q} \notag\\
&=\left(
b_{1}-\delta_{qq^{\prime}}b_{1}^{\prime}
\right)\int\mathrm{d}^{3}r\ \phi^{q^{\prime}*}_{\beta}\phi_{\alpha}^{q^{\prime}}\nabla\phi_{\mu}^{q*}\cdot\nabla\phi_{\nu}^{q},
\end{align}
\begin{equation}
\begin{split}
\label{eq:nucleon density QRPA}
\rho_{q}&=\sum_{\alpha\in q}v_{\alpha}^{2}|\phi_{\alpha}^{q}|^{2}\\
&+\eta\sum_{\substack{\alpha\beta\in q\\ \alpha\geq\beta}}\zeta^{+}_{\alpha\beta}(\phi^{q*}_{\beta}\phi_{\alpha}^{q}X_{\alpha\beta}^{q}+\phi^{q*}_{\alpha}\phi_{\beta}^{q}Y_{\alpha\beta}^{q})+O(\eta^{2}),
\end{split}
\end{equation}
where $b_{1}$ and $b_{1}^{\prime}$ are coefficients in the Skyrme force \cite{Maruhn:2013mpa}, and $\rho_{q}$ is the density of nucleon $q$. The symmetrical properties such as $X_{\alpha\beta}^{q}=X_{\bar{\beta}\bar{\alpha}}^{q}$, $Y_{\alpha\beta}^{q}=Y_{\bar{\beta}\bar{\alpha}}^{q}$, and $\phi^{q*}_{\bar{\alpha}}\phi_{\bar{\beta}}^{q}=\phi^{q*}_{\beta}\phi_{\alpha}^{q}$ are used to derive the second line of Eq.~(\ref{eq:nucleon density QRPA}). As implied in Eq.~(\ref{eq:derivative of effective mass}), integral terms in the QRPA matrices can be calculated by multiplying $\zeta^{\pm}_{\mu\nu}\zeta^{\pm}_{\alpha\beta}$ with the residual interaction derived in the same way as RPA calculation (e.g. Eqs.~(32) and (39) in Ref.~\cite{Sasaki:2022ipn}). The $\rho_{q}$ in Eq.~(\ref{eq:nucleon density QRPA}) contains a factor $\zeta^{+}_{\alpha\beta}$ in the linear term of $\eta$ and such a property is also confirmed in other time-even fields such as the spin-orbit density $\vec{J}_{q}$, and the kinetic energy density $\tau_{q}$ \cite{Maruhn:2013mpa}. On the other hand, the time-odd fields such as the current density $\vec{j}_{q}$ and the spin density $\vec{s}_{q}$ \cite{Maruhn:2013mpa} include a factor $\zeta^{-}_{\alpha\beta}$ in $O(\eta)$ (e.g. see Eq.~(\ref{eq:spin density})). $\zeta^{+}_{\alpha\beta}$ and $\zeta^{-}_{\alpha\beta}$ in Eqs.~(\ref{eq:QRPA matrix A}) and (\ref{eq:QRPA matrix B}) originated from the linear terms of $\eta$ in time-even and time-odd fields.

The frequency $\omega$ in Eq.~(\ref{eq:QRPAeq}) is decomposed into real and imaginary parts: $\omega=E+i\gamma/2$, where $E$ is the incoming photon energy and $\gamma$ is the Lorentzian width. The signs of $\tau$ in Eqs.~(\ref{eq:F20}) and (\ref{eq:F02}) are negative (positive) for the M1 (E1) operator due to the time-odd (even) operator. 
 
The forward and backward amplitudes are used to calculate the transition strength. In our QRPA, the transition strength of Eq.~(\ref{eq:transition strength QRPA}) is described by
 \begin{equation}
 \frac{\mathrm{d}B(\omega;F)}{\mathrm{d}\omega}=-\frac{1}{\pi}\mathrm{Im}\sum_{\substack{\mu\nu\in q\\ \mu\geq\nu}}\zeta^{\tau}_{\mu\nu}(f_{\mu\nu}^{q*}X_{\mu\nu}^{q}+f_{\nu\mu}^{q*}Y_{\mu\nu}^{q}),\label{eq:transition strength QRPA HF+BCS}
 \end{equation}
 where we use Eqs.~(\ref{eq:F20}),(\ref{eq:F02}), and (\ref{eq:xytrans}). Then, the photoabsorption cross section of the M1 transition is given by \cite{Sasaki:2022ipn},
\begin{eqnarray}
\sigma_{\mathrm{abs}}(E;\mathrm{M}1)&=&\frac{16\pi^{3}}{9\hbar c}E\sum_{K=0,\pm1}\frac{\mathrm{d}B(\omega;M_{K})}{\mathrm{d}\omega}, \label{eq:cross section M1}
\end{eqnarray}
where $M_{K}$ is the M1 operator written as,
\begin{equation}
\begin{split}
\label{eq:magnetic dipole operator}
M_{K}=\mu_{N}\sum_{i=1}^{A}\left(
g_{s}^{(i)}\frac{\vec{\sigma}_{i}}{2}+g_{l}^{(i)}\vec{l}_{i}
\right)\cdot\vec{\nabla}\left(
r_{i}Y_{1K}(\theta_{i},\varphi_{i})
\right),
\end{split}
\end{equation}
where the $(r_{i},\theta_{i},\varphi_{i})$ is the spherical coordinate of nucleon $i$, and $g_{l}^{(i)} = 0(1)$ and $g_{s}^{(i)}=-3.826(5.586)$ for neutrons (protons). The photoabsorption cross section of the E1 transition is calculated in the same way as Eq.~(\ref{eq:cross section M1}) with the E1 operator $D_{K}(K=0,\pm1)$ \cite{Sasaki:2022ipn} instead of $M_{K}$.

\subsection{Elimination of spurious modes}
\label{sec:spurious}

The (Q)RPA theory has spurious modes at zero energy ($\omega=0$) corresponding to the collective motion of the whole nucleus associated with the violation of symmetries such as translational and rotational symmetries in the intrinsic Hamiltonian \cite{ring2004nuclear}. In (Q)RPA calculations, such a spurious mode usually appears at low energy and couples with the physical states due to the discretized coordinate space and the limited size of the configuration space \cite{Stetcu:2002zf,Nakatsukasa:2007qj,Repko:2018gcn}. The admixture of the spurious mode can be eliminated with the symmetry operator $P$ and the conjugate operator $Q$ satisfying $\left[Q,P \right]=i\hbar$. In order to remove the spurious mode, we renormalize the external field $F$ with a description similar to Ref.~\cite{Repko:2018gcn},
\begin{equation}
\label{eq:elimination}
\tilde{F}=F-\frac{i}{\hbar}\bra{0}\left[
P,F
\right]\ket{0}Q+\frac{i}{\hbar}\bra{0}\left[
Q,F
\right]\ket{0}P,
\end{equation}
\begin{eqnarray}
\label{eq:symmetry operator}
    Q^{20}_{\mu\nu}=-(Q_{0})^{q}_{\mu\bar{\nu}},\ 
    P^{20}_{\mu\nu}=-(P_{0})^{q}_{\mu\bar{\nu}},
\end{eqnarray}
\begin{eqnarray}
\left(
\begin{array}{cc}
     A & B \\
     B^{*} & A^{*}
\end{array}
\right)
\left(
\begin{array}{c}
     Q_{0}\\
     -Q^{*}_{0}
\end{array}
\right)
=-\frac{i\hbar}{M_{0}}
\left(
\begin{array}{c}
     P_{0}\\
     P^{*}_{0}
\end{array}
\right),
\end{eqnarray}
\begin{eqnarray}
\label{eq:inertia}
M_{0}&=&2\left\{
\mathrm{Re}(P_{0})(A+B)^{-1}\mathrm{Re}(P_{0})\right.\nonumber\\
&&  \left.+\mathrm{Im}(P_{0})(A-B)^{-1}\mathrm{Im}(P_{0})
\right\},
\end{eqnarray}
where 
$M_{0}$ is the inertia for the spurious mode, and $\ket{0}$ is the QRPA vacuum approximated by the HF+BCS ground state within the quasiboson approximation. For the M1 transition with $F=M_{\pm1}$, the spurious mode appears at low energy due to the violation of the rotational symmetry for axially deformed nuclei and the spurious mode can be eliminated with the total angular momentum operator, $P=\sum_{i=1}^{A}\mp\frac{1}{\sqrt{2}}(J_{x}\pm iJ_{y})_{i}$ in Eqs.~(\ref{eq:elimination})--(\ref{eq:inertia}). For the E1 transition, the translation of the center-of-mass induces the spurious modes, and Eqs.~(\ref{eq:elimination})--(\ref{eq:inertia}) are used to eliminate them with the total momentum operator, $P=\sum_{i=1}^{A}(-i\hbar)(\nabla_{K})_{i} (K=0,\pm1)$. We remark that such an elimination for the E1 transition is equivalent to imposing an effective charge of a neutron, $e_{\mathrm{eff}}^{(n)}=-eZ/A$, and that of a proton, $e_{\mathrm{eff}}^{(p)}=eN/A$ on the E1 operator \cite{Repko:2018gcn}.

\subsection{Microscopic calculations for neutron capture}
\label{sec:capture cross sections}

We briefly review the calculation of neutron capture reactions following Ref.~\cite{Mumpower:2017gqj}. The photoabsorption cross sections for E1 and M1 transitions are used to calculate the neutron capture cross section based on the statistical Hauser-Feshbach model with the width fluctuation correction. In this statistical model, the formula for the radiative capture process, where a neutron and lumped $\gamma$-ray channels are involved, is written as,
\begin{equation}
\label{eq:capture cross section}
    \sigma_{n\gamma}(E_{n})=\frac{\pi}{k_{n}^{2}}\sum_{J\Pi}g_{c}\frac{T_{n}^{J\Pi}T_{\gamma}^{J\Pi}}{T_{n}^{J\Pi}+T_{\gamma}^{J\Pi}}W_{n\gamma}^{J\Pi},
\end{equation}
where $E_{n}$ is the incident neutron energy, $k_{n}$ is the incident neutron wave number, $g_{c}$ is the spin statistical factor, $W_{n\gamma}$ is the width fluctuation correction factor \cite{Kawano2015prc}, $T_{\gamma}$ is the lumped $\gamma$-ray transmission coefficient, and $T_{n}$ is the neutron transmission coefficient. The indices $J$ and $\Pi$ in the sum are the possible spin and the parity of the compound state.\\
\indent The lumped $\gamma$-ray transmission coefficient is given by
\begin{equation}
\label{eq:transmission gamma}
    T_{\gamma}^{J\Pi}=\sum_{j^{\pi} XL}\int^{E_{0}}_{0}\mathrm{d}E_{x}2\pi E_{\gamma}^{2L+1}f_{XL}(E_{\gamma})\rho(E_{x},j^{\pi}),
\end{equation}
where $E_{0}=E_{n}+S_{n}$ is the total excitation energy, $S_{n}$ is the neutron separation energy of the target nucleus, $E_{\gamma}$ is the emitted photon energy, $j^{\pi}$ is the spin and parity of the final state after the $\gamma$-decay, $E_{x}=E_{0}-E_{\gamma}$ is the excitation energy of the final state, $\rho$ is the level density at $E_x$, and $f_{XL}$ is the $\gamma$-ray strength function of the type of the transition $X(=E, M)$ and multipolarity $L$. The $\gamma$-ray strength function can be expressed in terms of the photoabsorption cross sections,
\begin{eqnarray}
\label{eq:gamma strength function}
f_{XL}(E_{\gamma})=\frac{\sigma_{abs}(E_{\gamma};XL)}{3\pi(\hbar c)^{2}E_{\gamma}},
\end{eqnarray}
where $\sigma_{abs}(E_{\gamma};XL)$ is the photoabsorption cross section for $XL$ transition. The above relation enables the application of the photoabsorption cross sections of QRPA to the microscopic calculation for the capture cross section without any experimental data of giant resonances.


\section{Results and Discussions}
\label{sec:Results and Discussions}

\subsection{M1 transition}
\label{sec:M1 transition QRPA}
We solve the QRPA equation with the M1 operator, $M_{K}(K=0,\pm1)$ as the external field and calculate the transition strength and the photoabsorption cross section of M1 transition for $^{156}\rm{Gd}$ following the description in Sec. \ref{sec:QRPA equation}. The single-particle states of HF+BCS are calculated as in the same setup in Ref.~\cite{Bonneau:2007dc}, and the QRPA matrices are calculated as in Ref.~\cite{Sasaki:2022ipn} employing the Skyrme parameters of SLy4 \cite{Chabanat:1997un}. Here, we consider the contribution from spin terms in the Skyrme force neglected in Ref.~\cite{Sasaki:2022ipn} that affects the spin-flip parts of the M1 transition \cite{Vesely:2009eb}. Such spin terms are involved in the residual interaction from the time-odd Hamiltonian, $h^{\mathrm{odd}}_{q}$ in Eqs.(\ref{eq:QRPA matrix A}) and (\ref{eq:QRPA matrix B}), and the detailed description is shown in Appendix \ref{sec:spin terms}. 

The QRPA equation in Eq.~(\ref{eq:QRPAeq}) is solved from $E=$62.5 keV to 20 MeV at every 62.5 keV with a fixed Lorentzian width $\gamma=125$ keV. With the symmetry of indices in the forward and backward amplitudes, we impose an asymmetry in the pairs of quasiparticles, $\mu\geq\nu$ and restrict the size of the configuration space as $u_{\mu}^{2},v_{\nu}^{2}>10^{-2}$, and $E_{\mu}+E_{\nu}<E_{\rm{cut}}=50\ \mathrm{MeV}$. The M1 spurious mode appears at lower incoming photon energy even though $E_{\rm cut}$ is relatively high. However, the transition strength of the spurious mode is negligible due to the elimination of Eq.~(\ref{eq:elimination}).

\begin{figure}[t]
\includegraphics[width=1\linewidth]{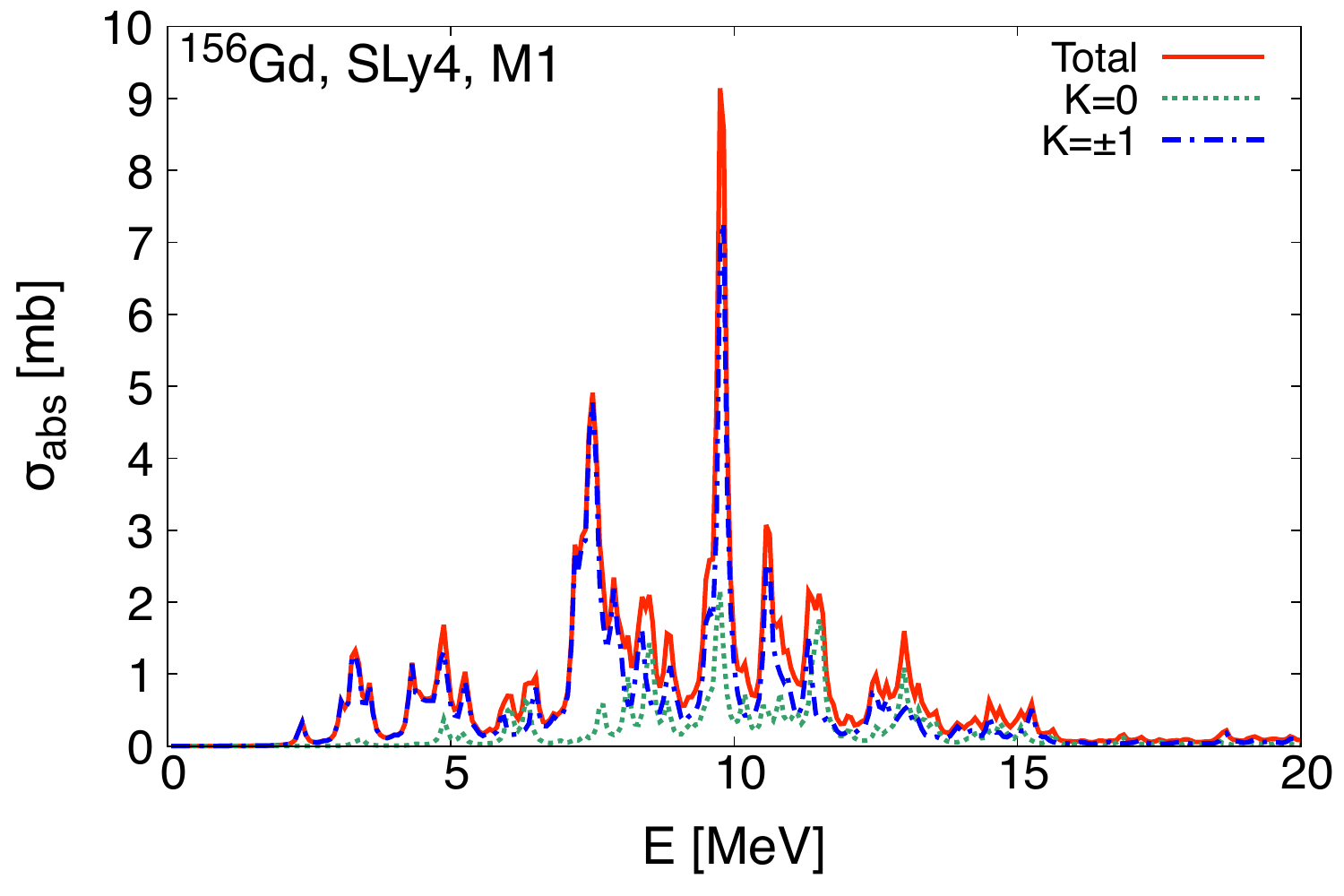}
\caption{
The calculated photoabsorption cross sections of the M1 transitions for $^{156}\rm{Gd}$. The solid line shows the total photoabsorption cross section in Eq.~(\ref{eq:cross section M1}). The dotted line is the component of $\mathrm{d}B(\omega,M_{0})/\mathrm{d}\omega$, and the dot-dashed line is for $\sum_{K=\pm1}\mathrm{d}B(\omega,M_{K})/\mathrm{d}\omega$.
}
\label{fig:QRPA cross sections M1 Gd156}
\end{figure}

Figure~\ref{fig:QRPA cross sections M1 Gd156} shows the calculated photoabsorption cross section in Eq.~(\ref{eq:cross section M1}) for $^{156}\mathrm{Gd}$ and the contributions from different values of $K$. The transition strength depends on the value of $K$ due to the deformation of $^{156}\mathrm{Gd}$. The excitation at low energy that can affect the neutron capture cross section mainly comes from the $K=\pm1$ mode. Without the contribution from Eq.~(\ref{eq:residual spin terms}), the result is similar to the case without the residual interaction as in the case of double magic nuclei \cite{Sasaki:2022ipn}. Such spin terms move the spin-flip strength to higher energies and separate contributions from the orbital and spin parts of Eq.~(\ref{eq:magnetic dipole operator}). The strong peaks at 7.5 MeV and 9.8 MeV in Fig.~\ref{fig:QRPA cross sections M1 Gd156} would reflect the double-humped structure as often found in heavy deformed nuclei \cite{RICHTER1995261}. 


\begin{figure}[t]
\includegraphics[width=1\linewidth]{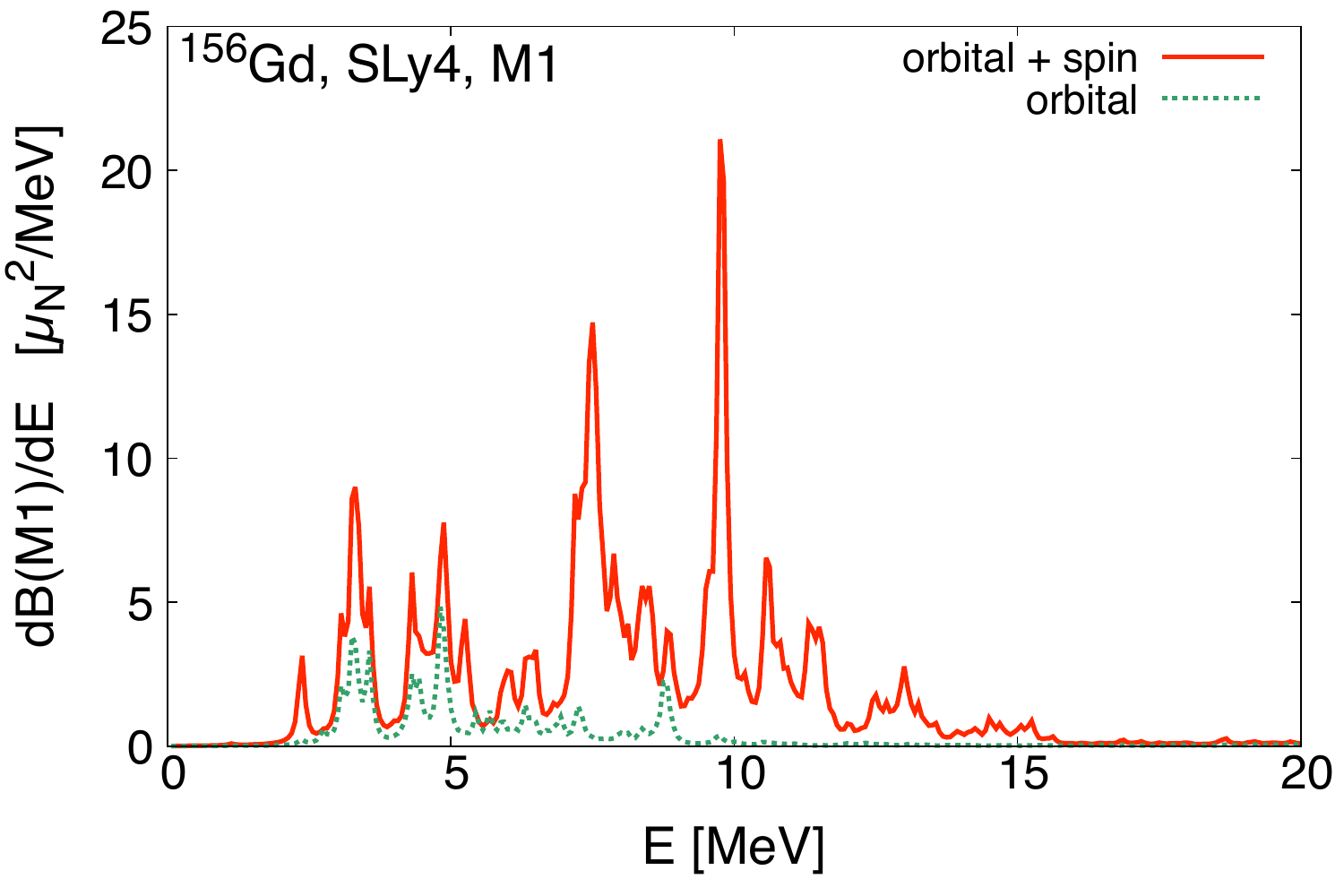}
\caption{
The strength functions of the M1 transitions for $^{156}\rm{Gd}$ with (solid line) and without (dotted line) the spin $g$ factor $g_{s}$ in Eq.~(\ref{eq:magnetic dipole operator}).
}
\label{fig:orbit spin Gd156}
\end{figure}

The transition strength is usually used to compare calculations with experimental data. 
In our QRPA calculation, the M1 transition strength is given by
\begin{equation}
\label{eq:transition strength distribution M1}
    \frac{\mathrm{d}B(\mathrm{M}1)}{\mathrm{d}E}=\sum_{K=0,\pm1}\frac{\mathrm{d}B(\omega;M_{K})}{\mathrm{d}\omega}.
\end{equation}
Figure~\ref{fig:orbit spin Gd156} shows the results of Eq.~(\ref{eq:transition strength distribution M1}) with and without the spin $g$ factor $g_{s}^{(i)}$ in Eq.~(\ref{eq:magnetic dipole operator}). It is clearly seen that the large strength in the 5 to 10 MeV range for orbital$+$spin (solid line) originated from the spin-flip M1 transition due to the finite value of $g_{s}^{(i)}$. The residual interaction induces the fragmentation of the spin-flip transitions and upshifts them up to about 15 MeV. From the energy integration of Eq.~(\ref{eq:transition strength distribution M1}), the total transition strength from $E=5$ MeV to 15 MeV is $\sum B(M1)=29.4\mu_{N}^{2}$, which is larger than the typical value ($\sim11\mu_{N}^{2}$) for heavy deformed nuclei \cite{RICHTER1995261}. Such overestimation of the total M1 transition strength is also reported by other published (Q)RPA calculations, and quenching of the spin $g$ factor was proposed to improve the agreement with experimental data \cite{harakeh2001giant}. By applying a typical value of the quenching factor, $g_{s,\mathrm{eff}}^{(i)}/g_{s}^{(i)}=0.6-0.7$  \cite{Kruzic:2020oqf}, the calculated $\sum B(M1)$ is reduced by a factor of 0.36 -- 0.49, since the contribution from the spin-flip transition is proportional to the square of $g_{s}^{(i)}$ in Eq.~(\ref{eq:magnetic dipole operator}). 

As shown in the dotted line of Fig.~\ref{fig:orbit spin Gd156}, the contribution from orbital motion is not negligible at low energies. The transition strength near 3.3 MeV can be seen as the M1 scissors mode in a macroscopic view of the collective motion. For rare-earth nuclei, the scissors mode appears around 3 MeV, and the total transition strength is reported to be $\sum B(M1)\sim 3\mu_{N}^{2}$ \cite{RICHTER1995261}. In our QRPA calculation, we obtain $\sum B(M1)=4.9\mu_{N}^{2}$ by integrating Eq.~(\ref{eq:transition strength distribution M1}) up to $4$ MeV. The difference between the solid and dotted lines in Fig.~\ref{fig:orbit spin Gd156} implies that the contribution of spin part in Eq.~(\ref{eq:magnetic dipole operator}) persists even at low energies. This suggests the overestimation of our calculated $\sum B(M1)$ might be reconciled by introducing the quench of $g_{s}^{(i)}$.

\begin{figure}[t]
\includegraphics[width=1\linewidth]{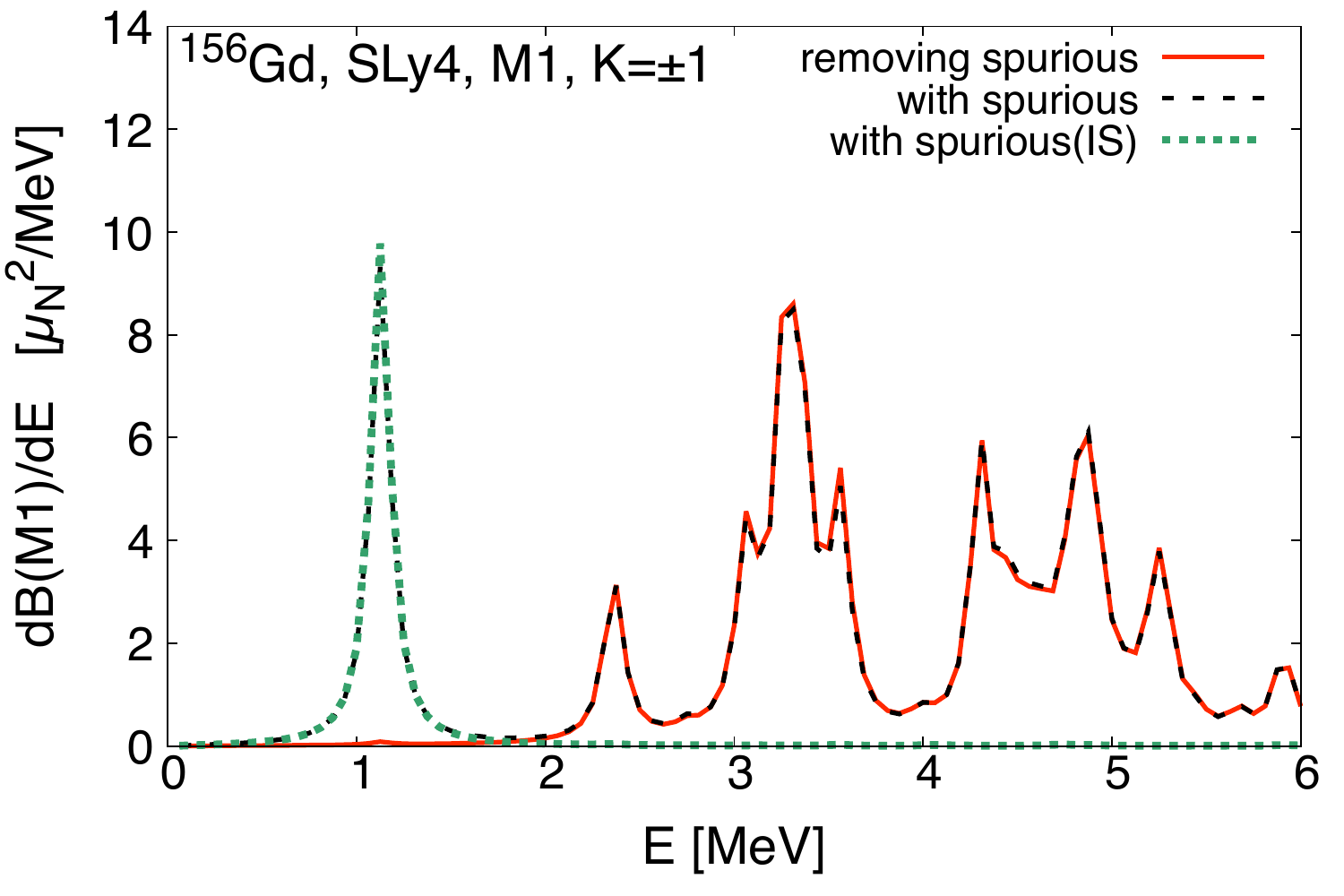}
\caption{
The elimination of the spurious mode from the M1 scissors mode with Eq.~(\ref{eq:elimination}). The solid line (dashed line) shows the result of $\sum_{K=\pm1}\mathrm{d}B(\omega,M_{K})/\mathrm{d}\omega$ in the case with (without) removing the spurious mode. The dotted line shows the contribution from the isoscalar operator in Eq.~(\ref{eq:magnetic moment isoscalar}) to the dashed line.
}
\label{fig:spurious Gd156}
\end{figure}

\subsection{Spurious mode from M1 isoscalar operator}
We discuss the role of elimination of the spurious mode following Sec.~\ref{sec:spurious}. In order to clarify the origin of the spurious mode, we separate the magnetic moments of nucleons in Eq.~(\ref{eq:magnetic dipole operator}) by two parts such as the isoscalar (IS) and isovector (IV) operators \cite{RICHTER199099},
\begin{eqnarray}
\label{eq:magnetic moment}
\vec{\mu}=\mu_{N}\sum_{i=1}^{A}\left(
g_{s}^{(i)}\frac{\vec{\sigma}_{i}}{2}+g_{l}^{(i)}\vec{l}_{i}
\right)=\mu_{N}\left(
\vec{\mu}_{\mathrm{IS}}+\vec{\mu}_{\mathrm{IV}}
\right),
\end{eqnarray}

\begin{eqnarray}
\label{eq:magnetic moment isoscalar}
\vec{\mu}_{\mathrm{IS}}=\frac{1}{2}\vec{J}+\frac{g^{p}_{s}+g^{n}_{s}-1}{2}\sum_{i=1}^{A}
\frac{\vec{\sigma}_{i}}{2},
\end{eqnarray}

\begin{eqnarray}
\label{eq:magnetic moment isovector}
\vec{\mu}_{\mathrm{IV}}=-\sum_{i=1}^{A}\tau_{zi}\vec{l}_{i}
+\left(
g^{p}_{s}-g^{n}_{s}
\right)\sum_{i=1}^{A}\tau_{zi}\frac{\vec{\sigma}_{i}}{2},
\end{eqnarray}
where $\vec{J}=\sum_{i=1}^{A}(\vec{l}_{i}+\vec{\sigma}_{i}/2)$ is the total angular momentum operator and $\tau_{zi}$ is $1/2 (-1/2)$ for neutrons (protons). 

The solid line (dashed line) in Fig.~\ref{fig:spurious Gd156} shows the result of $\sum_{K=\pm1}\mathrm{d}B(\omega,M_{K})/\mathrm{d}\omega$ with (without) removing the spurious mode. By comparing these two lines, we can see the spurious mode at 1.2 MeV in the dashed line. Such M1 spurious mode at low energy is consistent with the result in Ref.~\cite{Repko:2018gcn}. The dotted line in Fig.~\ref{fig:spurious Gd156} shows the result of $\sum_{K=\pm1}\mathrm{d}B(\omega,M_{K})/\mathrm{d}\omega$ using only Eq.~(\ref{eq:magnetic moment isoscalar}) without the elimination of the spurious mode. The dotted line is almost equivalent to the dashed line in $E<2$ MeV and negligible at higher energy. Therefore, the IS mode gives rise to the spurious mode and hardly affects physical excitations. The spurious mode of the M1 transition is associated with the collective rotation of the whole nucleus around an axis perpendicular to the symmetry axis of the axially deformed nucleus \cite{Repko:2018gcn} and $\vec{J}$ in Eq.~(\ref{eq:magnetic moment isoscalar}) dominantly induces such a collective rotation. The contribution from the spin term in the IS mode is negligible due to the opposite signs of $g_{s}^{p}$ and $g_{s}^{n}$. Such results of QRPA follow up a qualitative nature of the M1 transition as discussed in Ref.~\cite{Heyde:2010ng}.


\subsection{Neutron capture reactions}


We calculate the neutron capture cross sections for Gd isotopes following the discussion in Sec.~\ref{sec:capture cross sections}. We use the coupled-channels Hauser-Feshbach code $\mathrm{CoH}_{3}$ \cite{Kawano2021} with the $\gamma$-ray strength functions of various $XL$. As a default setting, $\mathrm{CoH}_{3}$ employs the standard Lorentzian profiles for the M1, E2, M2, and E3 transitions and the generalized Lorentzian form \cite{Kopecky:1990PRC} for the E1 transition. The large contribution to the $\gamma$-ray strength function comes from the E1 and M1 transitions. Here, we calculate both the E1 and M1 transitions for Gd isotopes with the same numerical setup as in Sec.~\ref{sec:M1 transition QRPA} and then apply the QRPA results to Eq.~(\ref{eq:gamma strength function}) to calculate the capture cross section instead of the $\mathrm{CoH}_{3}$ internal strength functions. Note that photoabsorption cross sections of an even-odd nucleus, $(Z,A+1)$ are approximated to those of an even-even nucleus, $(Z,A)$ when we calculate Eq.~(\ref{eq:capture cross section}) of $(Z,A)$ because the photoabsorption cross section varies weakly as the target mass number. 

\begin{figure}[t]
\includegraphics[width=1\linewidth]{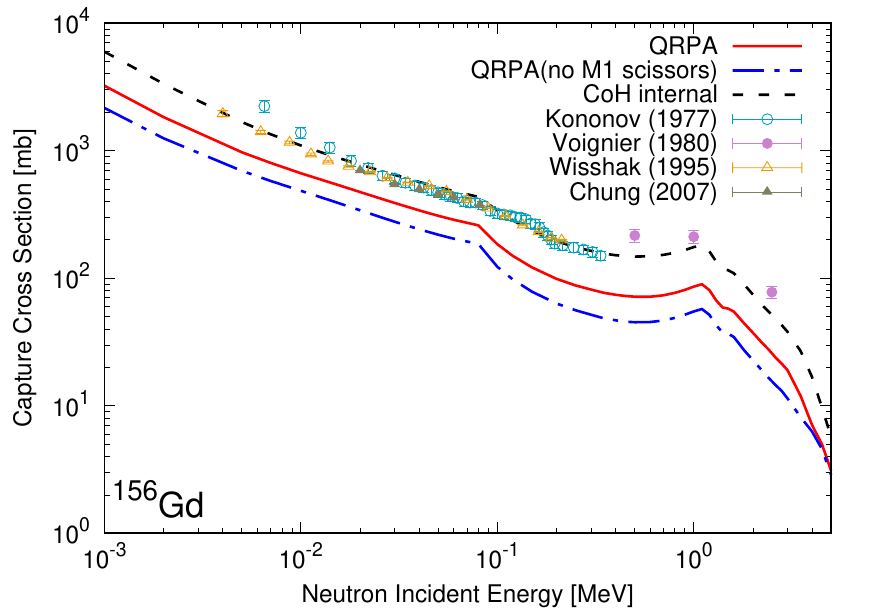}
\caption{
The neutron capture cross sections on $^{156}\rm{Gd}$. The solid line shows the result of Eq.~(\ref{eq:capture cross section}) with photoabsorption cross sections of the E1 and M1 transitions in QRPA. The dash-dotted line shows the result with QRPA cross sections neglecting the contribution from the M1 scissors mode ($E_{\gamma}\leq4$ MeV) in Eq.~(\ref{eq:gamma strength function}) with $XL=$M1. The dashed line shows the result with $\mathrm{CoH}_{3}$ internal strength functions. The symbols are available experimental data \cite{Kononov:1977capture,Voignier:1981capture,Wisshak:1995capture,Chung:2007capture}.
}
\label{fig:capture Gd156}
\end{figure}

\begin{figure}[htbp]
\includegraphics[width=1\linewidth]{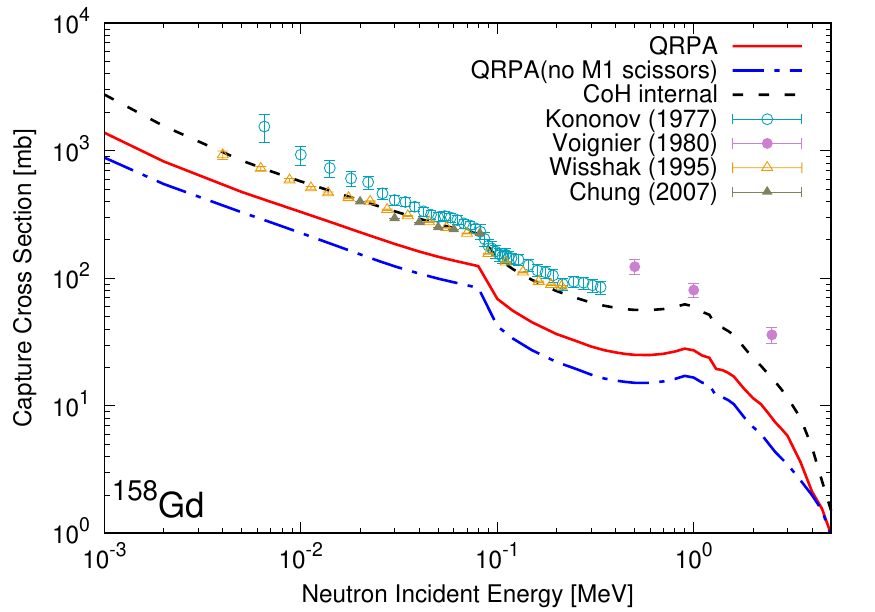}
\caption{
Same as in Fig.~\ref{fig:capture Gd156}, but for $^{158}\rm{Gd}$.
}
\label{fig:capture reacrions Gd isotopes}
\end{figure}

\begin{figure}[t]
\includegraphics[width=1\linewidth]{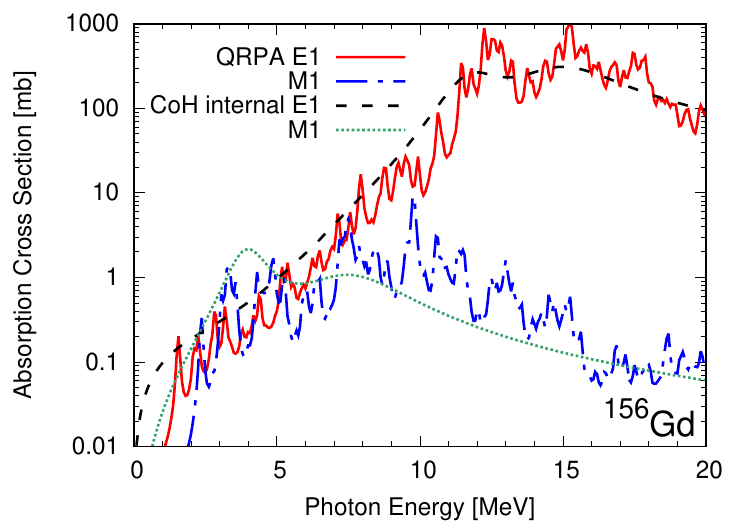}
\caption{
The photoabsorption cross sections of E1 and M1 transitions for $^{156}\rm{Gd}$, which are employed to calculate the neutron capture cross sections shown by the solid and dashed lines in Fig.~\ref{fig:capture Gd156}.
}
\label{fig:photoabs for capture Gd156}
\end{figure}

Figure~\ref{fig:capture Gd156} shows the calculated neutron capture cross sections on $^{156}\mathrm{Gd}$ compared with experimental data. By comparing the solid and dash-dotted lines, we can see an enhancement of the calculated capture cross section caused by the M1 scissors mode in the low energy region ($E_{\gamma}\leq4$ MeV) as in Ref.~\cite{Mumpower:2017gqj}. The calculated capture cross section is sensitive to the strengths of the excitations at a few MeV of $E_{\gamma}$ because the level density of the compound state $\rho$ in Eq.~(\ref{eq:transmission gamma}) increases with the excitation energy, $E_{x}=E_{0}-E_{\gamma}$. The shapes of the two lines are similar in a wide range of neutron incident energy, and the difference in magnitude is characterized by the average $\gamma$-ray width $\average{\Gamma_{\gamma}}$ calculated with the transmission coefficient in Eq.~(\ref{eq:transmission gamma}) and the average $s$-wave neutron level spacing $D_{0}$ \cite{Mumpower:2017gqj}. The value of the $\average{\Gamma_{\gamma}}$ for the solid line (dash-dotted line) in Fig.~\ref{fig:capture Gd156} is 0.031 eV (0.019 eV), which indicates that the contribution from the scissors mode to the calculated capture cross section is $1-{0.019}/{0.031}\approx39\%$. Such a significant impact of the low energy M1 transition was also found in Refs.~\cite{Mumpower:2017gqj,Goriely:2018gig} and produces results in better agreement with the experimental data.

The calculated neutron capture cross section with the QRPA photoabsorption cross section is systematically lower than the experimental data, as shown in Fig.~\ref{fig:capture Gd156} by the solid line. For a quantitative discussion, the dashed line is the calculated capture cross section with the $\mathrm{CoH}_{3}$ internal strength functions, which are globally parameterized giant resonances adjusted to available experimental data. The average  $\gamma$-ray width of $\mathrm{CoH}_{3}$ internal is $\average{\Gamma_{\gamma}}=0.067$ eV so our QRPA result underestimates the capture cross section by $\frac{0.031}{0.067}\approx 46\%$.

Although we demonstrated this for $^{156}\mathrm{Gd}$, almost the same properties are confirmed for other Gd isotopes. For example, Figure~\ref{fig:capture reacrions Gd isotopes} shows neutron capture cross sections on $^{158}\mathrm{Gd}$. The shapes of the calculated results are similar over a wide range of neutron incident energy, but the magnitudes are different. The values of $\average{\Gamma_{\gamma}}$ for QRPA (solid line), QRPA without the scissors mode (dash-dotted line), and $\mathrm{CoH}_{3}$ internal (dashed line) are 0.031 eV, 0.018 eV, and 0.073 eV, respectively. Therefore, the scissors mode contributes to about $42\%$ of the QRPA capture cross section, and the QRPA result is about $42\%$ of the experimental data.

Figure ~\ref{fig:photoabs for capture Gd156} shows the employed E1 and M1 cross sections for both QRPA (solid line) and $\mathrm{CoH}_{3}$ internal (dashed line) in Fig.~\ref{fig:capture Gd156}. The QRPA result of the E1 transition (solid line) in Fig.~\ref{fig:photoabs for capture Gd156} reproduces well the split of GDR peaks for deformed nuclei as in Ref.~\cite{Sasaki:2022ipn}. The tail on the lower side of GDR dominantly contributes to Eq.~(\ref{eq:transmission gamma}) due to the large $\rho(E_{x},j^{\pi})$, and the solid line is smaller than the dashed line at small photon energy, which results in the underestimation of the QRPA result in Fig.~\ref{fig:capture Gd156}. Here, we emphasis that the underestimation problem can hardly be resolved by the theoretical improvement of the microscopic M1 calculation alone, because, as discussed in Sec.~\ref{sec:M1 transition QRPA}, our QRPA calculation overestimates the value of $\sum B(M1)$ for the scissors mode so a more realistic calculation is supposed to reduce the values of both the $\sum B(M1)$ and $\average{\Gamma_{\gamma}}$, which is against the situation of neutron capture cases. We may envisage uncertainties in the E1 transition at low energies also impact the calculated capture cross sections. The E1 transition strength in QRPA should be enhanced by 3--5 times to reproduce the calculated capture cross section of $\mathrm{CoH}_{3}$ internal. We used a fixed Lorentzian width, $\gamma=125$ keV for our QRPA calculation, but the energy and temperature-dependent width \cite{Kopecky:1990PRC} may affect the low energy tail of the GDR. The toroidal dipole resonance is predicted in the same energy region of the pygmy resonance \cite{Repko:2019nej}, and such a low energy E1 excitation can be involved in our QRPA by considering the second-order terms of the E1 operator ignored in the long-wavelength limit \cite{Kvasil:2011yk}. Furthermore, the calculated capture cross section can be enhanced in the microscopic calculations considering the phonon coupling as in quasiparticle time blocking approximation (QTBA) \cite{Avdeenkov:2011zz} and quasiparticle-phonon model (QPM) \cite{Tsoneva:2015cra}.

\section{Conclusion}\label{sec:Conclusion}
We extended our noniterative FAM-RPA to the framework of FAM-QRPA with the HF+BCS single-particle states and solved the QRPA equation to study the M1 transition for deformed gadolinium isotopes. We showed large spin-flip transitions from 5 to 10 MeV and the orbital transition around 3 MeV where the M1 scissors mode was experimentally confirmed in the deformed rare-earth nuclei. We demonstrated that the spurious mode of the M1 transition originated from the IS part of the M1 operator. Although our result overestimates the total M1 transition strength, it can be reduced when we consider quenching of the spin $g$-factor as proposed in previous QRPA studies. 

Finally we applied the QRPA results of E1 and M1 transitions to calculations of neutron capture reactions based on the statistical Hauser-Feshbach theory. The contribution from the low energy M1 transition to the total calculated capture cross section is about $40\%$, and our calculation underestimates the capture cross section by $40-50\%$ compared with the experimental data. Improvements in the cross section would probably be possible by considering some uncertainty of the low energy E1 transition neglected in our QRPA calculation. This we leave for future work.

\begin{acknowledgments}
This work was partially support by the Office of Defense Nuclear Nonproliferation Research \& Development (DNN R\&D), National Nuclear Security Administration,
U.S. Department of Energy. This work was carried out under the
auspices of the National Nuclear Security Administration of the
U.S. Department of Energy at Los Alamos National Laboratory under
Contract No.~89233218CNA000001.
\end{acknowledgments}
\appendix

\section{Spin terms in the residual interaction}
\label{sec:spin terms}
In QRPA calculation, we consider the contribution from spin terms in the Skyrme forces ignored in our previous RPA calculation \cite{Sasaki:2022ipn}. The Skyrme energy density, $\mathcal{H}_{\mathrm{Sk}}$ includes the spin terms labeled with $\Tilde{b}_{i},\Tilde{b}_{i}^{\prime}(i=0,2,3)$ \cite{Vesely:2009eb},
\begin{eqnarray}
\label{eq:skyrme spin energy density}
\mathcal{E}_{\mathrm{spin}}&=&\frac{\tilde{b}_{0}}{2}|\vec{s}|^{2}-\frac{\tilde{b}_{0}^{\prime}}{2}\sum_{q}|\vec{s}_{q}|^{2}\nonumber\\
&+&\frac{\tilde{b}_{3}}{3}\rho^{\alpha}|\vec{s}|^{2}-\frac{\tilde{b}_{3}^{\prime}}{3}\rho^{\alpha}\sum_{q}|\vec{s}_{q}|^{2}\nonumber\\
&-&\frac{\tilde{b}_{2}}{2}\vec{s}\cdot\Delta\vec{s}+\frac{\tilde{b}_{2}^{\prime}}{2}\sum_{q}\vec{s}_{q}\cdot\Delta\vec{s}_{q},
\end{eqnarray}
where $\vec{s}=\vec{s}_{n}+\vec{s}_{p}$ is the summation of the spin density of nucleon $q(=n,p)$,
\begin{equation}
\vec{s}_{q}=\eta\sum_{\substack{\alpha\beta\in q\\ \alpha\geq\beta}}\zeta^{-}_{\alpha\beta}(\phi^{q*}_{\beta}\vec{\sigma}\phi_{\alpha}^{q}X_{\alpha\beta}^{q}+\phi^{q*}_{\alpha}\vec{\sigma}\phi_{\beta}^{q}Y_{\alpha\beta}^{q})+O(\eta^{2}),\label{eq:spin density}
\end{equation}
where $\phi^{q*}_{\bar{\alpha}}\vec{\sigma}\phi_{\bar{\beta}}^{q}=-\phi^{q*}_{\beta}\vec{\sigma}\phi_{\alpha}^{q}$ is used in the summation of $\alpha,\beta$. The $b$ coefficients in Eq.~(\ref{eq:skyrme spin energy density}) can be described in terms of $t$ and $x$ coefficients in the Skyrme forces \cite{Bender:2003jk}. 
The functional derivatives such as $\delta/\delta\vec{s}_{q}(\int\mathrm{d}^{3}r\mathcal{E}_{\mathrm{spin}})$ and $\delta/\delta\rho_{q}(\int\mathrm{d}^{3}r\mathcal{E}_{\mathrm{spin}})$ generally induce potentials in the single-particle Hamiltonian for static calculations \cite{Bender:2003jk}. In the case of even-even nuclei satisfying the time-reversal symmetry, the time-odd potentials derived from such functional derivatives do not affect the static HF+BCS calculation. However, in dynamical calculations like (Q)RPA, the second derivative $\delta^{2}/\delta\vec{s}_{q}\delta\vec{s}_{q^{\prime}}(\int\mathrm{d}^{3}r\mathcal{E}_{\mathrm{spin}})$ can induce a residual interaction in $\delta h^{\mathrm{odd}}$ \cite{Vesely:2009eb}. The contribution from Eq.~(\ref{eq:skyrme spin energy density}) to an integral term proportional to $\zeta^{-}_{\mu\nu}\zeta^{-}_{\alpha\beta}$ in Eq.~(\ref{eq:QRPA matrix A}) is written as
\begin{eqnarray}
\label{eq:residual spin terms}
&&(\tilde{b}_{0}-\delta_{qq^{\prime}}\tilde{b}_{0}^{\prime})\int\mathrm{d}^{3}r\ (\phi_{\mu}^{q*}\vec{\sigma}\phi_{\nu}^{q})\cdot(\phi_{\beta}^{q^{\prime}*}\vec{\sigma}\phi_{\alpha}^{q^{\prime}})\nonumber\\
&+&(\tilde{b}_{3}-\delta_{qq^{\prime}}\tilde{b}_{3}^{\prime})\int\mathrm{d}^{3}r\ \frac{2}{3}(\rho_{0})^{\alpha}(\phi_{\mu}^{q*}\vec{\sigma}\phi_{\nu}^{q})\cdot(\phi_{\beta}^{q^{\prime}*}\vec{\sigma}\phi_{\alpha}^{q^{\prime}})\nonumber\\
&+&(\tilde{b}_{2}-\delta_{qq^{\prime}}\tilde{b}_{2}^{\prime})\int\mathrm{d}^{3}r\ \nabla(\phi_{\mu}^{q*}\vec{\sigma}\phi_{\nu}^{q})\cdot\nabla(\phi_{\beta}^{q^{\prime}*}\vec{\sigma}\phi_{\alpha}^{q^{\prime}}),
\end{eqnarray}
where $\rho_{0}=(\rho_{n}+\rho_{p})_{\eta=0}$ in Eq.~(\ref{eq:nucleon density QRPA}). The effect on Eq.~(\ref{eq:QRPA matrix B}) is easily derived from an exchange, $(\phi_{\beta}^{q^{\prime}*}\vec{\sigma}\phi_{\alpha}^{q^{\prime}})\to(\phi_{\beta}^{q^{\prime}*}\vec{\sigma}\phi_{\alpha}^{q^{\prime}})^{*}$ in Eq.~(\ref{eq:residual spin terms}). In our QRPA calculation, $\phi^{q}_{\mu}$ is expanded in the cylindrical coordinate space as in Ref.~\cite{Vautherin:1973zz} and the above residual interaction is calculated in analogy with our previous RPA calculation \cite{Sasaki:2022ipn}.


\bibliography{ref}


\end{document}